# Second Screen User Profiling and Multi-level Smart Recommendations in the context of Social TVs

Angelos Valsamis, Alexandros Psychas, Fotis Aisopos, Andreas Menychtas and Theodora Varvarigou

Distributed, Knowledge and Media Systems Group, National Technical University of Athens, Greece

*angval@central.ntua.gr, {alps, fotais, ameny }@mail.ntua.gr, dora@telecom.ntua.gr*

**Abstract.** In the context of Social TV, the increasing popularity of first and second screen users, interacting and posting content online, illustrates new business opportunities and related technical challenges, in order to enrich user experience on such environments. SAM (Socializing Around Media) project uses Social Media-connected infrastructure to deal with the aforementioned challenges, providing intelligent user context management models and mechanisms capturing social patterns, to apply collaborative filtering techniques and personalized recommendations towards this direction. This paper presents the Context Management mechanism of SAM, running in a Social TV environment to provide smart recommendations regarding for first and second screen content. Work presented is evaluated using real movie rating dataset found online, to validate the SAM's approach in terms of effectiveness as well as efficiency.

## 1. Introduction

The usage of mobile devices has become one of the leading daily activities. This phenomenon extends to the usage of those devices in parallel with other devices also. SAM project [1] aims at exploiting, researching and creating the appropriate technologies that revolve around the usage of mobile devices simultaneously with TV, the so called $2^{nd}$ screen phenomenon[1]. The software created for the purposes of SAM revolves around the creation of a complete experience for the user delivered in to his mobile device during a TV program. In a very simple way users get multimedia content (in the form of widgets) about the TV program they are watching in to their mobile devices. This Content varies from simple information about the characters of the TV program, e-commerce material to social media content about the program.

Delivering multimedia content to the user mobile devices has many challenges and requirements. The creation of a mechanism that delivers personalized content as well as the contextualisation of this content are the requirements that drove the research and developments of this paper. The work is summarized into two main objectives:

- Creating sophisticated Data through contextualisation of the information and content
- Create recommendation system for personalised content delivery

There is a vast variety of recommendation algorithms to implement in a system. In this paper the research focus not on creating new algorithms but creating sophisticated data to use as input in these algorithms. More specifically representing relations between users and multimedia content in a more refined way and further more adapt this data to be used as input to the well-established and tested algorithms for recommendation.

The rest of the document is organized as follows: The current section introduces the Social TV Context Management and Recommendation concepts. Section 2 presents the related work regarding Second

---

[1] Second Screen Society: http://www.2ndscreensociety.com/

Screen Context Management and graph Recommendations, while Section 3 analyzes the SAM model and approach. In Section 4, experiments' configuration as well as their results are provided and finally Section 5 concludes the current paper and discusses the work to be done in the future.

## 2. Related Work

### 2.1. Second Screen Context Management

Social TV and Second Screen are now some of the most emerging technologies, used for eLearning, Political Surveys or just Social Networking purposes. Cezar et al. [2] analyzed the usages of the Second Screen in an Interactive Television Environment, to control, enrich, share, and transfer TV content. This work provided an initial market assessment in the areas of media creation and distribution and subjected its prototype implementation to test by a dozen groups of users in a social setting. Giglietto & Selva [3] applied a content analysis to a big dataset of Second Screen tweets during an entire TV season, in order to clarify the relationship between TV political talk shows and related comments on social media. This study points out the effects of celebritization of politics and confirms the coexistence of different and interlinked forms of participation (with political prevailing on audience participation). Elaborating on personalized experiences, Geerts et al. [4] investigated a second screen companion application, stimulating social interaction in the living room, offering more insight into how viewers are experiencing second such applications, and contrasted this with the perspective of producers and actual usage data.

### 2.2. Context-based graph analysis and recommendations

Although much work has been carried out concerning movie/TV programs, Second Screen and Social TV recommendations are quite immature. Context-based recommendations using graphs are evidently the more efficient, as SQL databases are now obsolete for big data analytics. Demovic et al. [5] presented a suchlike approach, saving movie data in a graph and using Graph Traversal Algorithms to efficiently address user preferences. This work uses explicit user "likes" for movies or genres, but does not collect any contextual or social data. When it comes to Social TV Platforms, authors in [6] and [7] highlight the concept of context management and analysis in the frame of social enabled content delivery to Second Screen devices. These papers present a novel solution for media context management in a Neo4j graph database, and provide the baseline context of the current work.

### 2.3. kNN and Collaborative Filtering TV Recommendations

Collaborative filtering techniques are commonly used for TV program recommendations. Authors in [8] use collaborative filtering for such recommendations, enhanced with singular value decomposition resulting into a low-dimension item-based filtering with promising accuracy. Andrade and Almeida [9] adopt the k Nearest Neighbors (k-NN) algorithm to implement a hybrid strategy that combines collaborative filtering with a content-based method for delivering TV recommendations to individual users. K-NN is also employed in [10] and [11] to implement personalized popular program recommendation systems for digital TV data clustered by k-means. Both works generate datasets of user profiles, to examine resulting recommendations in terms of accuracy as well as computation time.

## 3. SAM Context Management and recommendations

### 3.1. Context Management Database

As described in the 2.3 section, graph databases are the leading solution for the data analysis and by extend recommendations systems. All the data created or imported in SAM project are stored in a Neo4j Graph database in order to be further analysed and used for recommendation purposes.

The structure of the graph database is a very important factor on how the algorithms for the recommendation will be used and also be optimised. Graph databases contain two types of data in general, the nodes and the edges. Nodes represent entities such as persons and multimedia Assets, edges represent relationships between the entities (nodes).

As far as the SAM graph database is concerned there are three types of nodes:

- **Assets**: which represent any type of multimedia content
- **Persons**: which represent every user that interacts with the Assets
- **Keywords**: which represent words that describe Assets

There are several types of edges that describe the relationships between these nodes:

- Has-keyword is the relationship that connects the Assets with the Keywords that most appropriately describes them.
- Consume, like, dislike, comment, full-screen, dismiss and show-more are relationships that describe interactions between a user (Person) and an Asset.

The way all these entities and interactions are used to produce recommendations is explained in the following paragraphs.

### 3.2. User actions

SAM first and second screen listeners collect various user actions and store them in order to be able to later recommend videos and/or widgets. In particular, actions concerning videos include:

1. Commenting on a movie: a comment in Twitter/Facebook/SAM DC that is being processed by SAM's sentiment analysis service.
2. Consume a movie: action of selecting a movie and start watching.
3. Initiate Full screen: action of pressing the enlargement button in order to see the movie in full screen mode.
4. Like/Dislike a movie: action of pressing the like/dislike button under each movie.

Actions concerning widgets include:

1. Like/Dislike a widget: The actions of pressing the like/dislike button under each widget.
2. Dismiss a widget: The action of pressing the dismiss button in order to hide a widget.

3. Show more: The action of pressing the show-more button, located under each widget that enlarge the widget's size and adds more info.

SAM's sentiment analysis service [12],[13], which is used to identify sentiment on widget and movie comments, is also able to perceive the sentiment polarity of a comment in regards to the movie's keywords. For example, if the user commented "That movie was awesome. Jennifer Lawrence's acting was spot on!" the sentiment analysis service will generate a positive number for the comment in regard to the movie, and also a positive number for the movie's keyword "Lawrence". Thus, we also identify a (indirect):

1. Comment on keyword.

A basic part of the analysis of the graph is to apply some kind of "weights" to the lines connecting users and assets. Setting +1 and -1 as absolute values of relevancy and irrelevancy respectively, we apply those values to user-asset relations that explicitly show such a rating ("like" weights for +1, "dislike" weights for -1). On the other hand, comments on assets are saved along with their sentiment polarity and intensity (percentage of positivity or negativity), thus we can apply for positive comments a decimal weight, ranging from (0, +1] and for negative comments from [-1, 0). Zero value obviously expresses neutrality.

However, consuming or pressing 'Full Screen' on a root asset also indicates some interest by the user. The same applies for pressing 'show more" on a specific widget in second screen, while dismissing it before it automatically closes indicates lack of interest. To capture those implicit patterns, we need to make sure that they will not totally overlap the explicit ones already mentioned. For example, if a user has "liked" an asset, but on the other hand dismissed it early on, this implies a weaker "like" or "interest" relation. The approach that we follow to make sure the overall weight (sum of weights) is mainly defined by "likes" / "dislikes" and only partly affected by other interactions is to apply to the latest a weight of

$$w_i = \frac{p_i}{t-1}$$

where p = polarity indication (+1,-1) and t = number of interaction types for this asset type. In this case, if an explicit interaction weight $w_e$ is contradictory to implicit weights $w_i$, the overall weight,

$$W = w_e + \sum w_i$$

will still bare the (now normalized) "polarity" of $w_e$.

Based on this, the cumulated weights of different interactions with 1st and 2nd screen elements are summarized in Table 1.

Table 1. Polarity contribution of the various user interactions.

|  | Movie | Widget | Keyword |
| --- | --- | --- | --- |
| Comment | (-1,1) | - | (-1,1) |
| Like | 1 | 1 |  |
| Dislike | -1 | -1 |  |
| Full screen | ¼ |  |  |
| Consume | ¼ |  |  |

| | |
|---|---|
| Dismiss | -⅓ |
| Show More | ⅓ |

### 3.3. Asset recommendations

When performing asset recommendations we identify two cases:
1. An interaction between the user and a widget/keyword that belongs to a movie. (Fig 1.)
2. An interaction between the user and a movie that shares widgets/keywords with another movie. (Fig 2.)

When we are asked to recommend movies, we use interactions with widgets and keywords (case 1) to calculate relevance with connected movies, not yet consumed by the user. We use interactions with movies (case 2) that share widgets and keywords with other movies, not yet consumed by the user.

In cases, for example, that a user has "liked" or commented positively for all widgets or keywords of a root asset (which may also exist in other videos as well), a strong indication of relevance to this root asset also exists. Similarly to the previous logic, we need to make sure that indirect relations to assets will not overlap a direct weight to it. Thus, for every rating to a connected widget/keyword we apply a weight of

$$W_w = \frac{r_x}{a + k + 1}$$

where $r_x$ = rating of neighboring node, $a$ = number of neighboring assets and $k$ = number of keywords connected to the "under investigation" asset.

In cases, where a user has "liked" a root asset, which shares keywords or widgets with the under investigation asset, a *weaker* indication of relevance has to be taken into account. Thus, for every rating to a movie connected with shared keyword/widget we apply a weight of

$$W_a = \frac{r_x}{(a + k + 1)2}$$

Therefore, the overall relevance weight of a person for an unconsumed asset now becomes:

$$W = W_w + W_a = \sum \frac{r_x}{a + k + 1} + \sum \frac{r_x}{(a + k + 1)2}$$

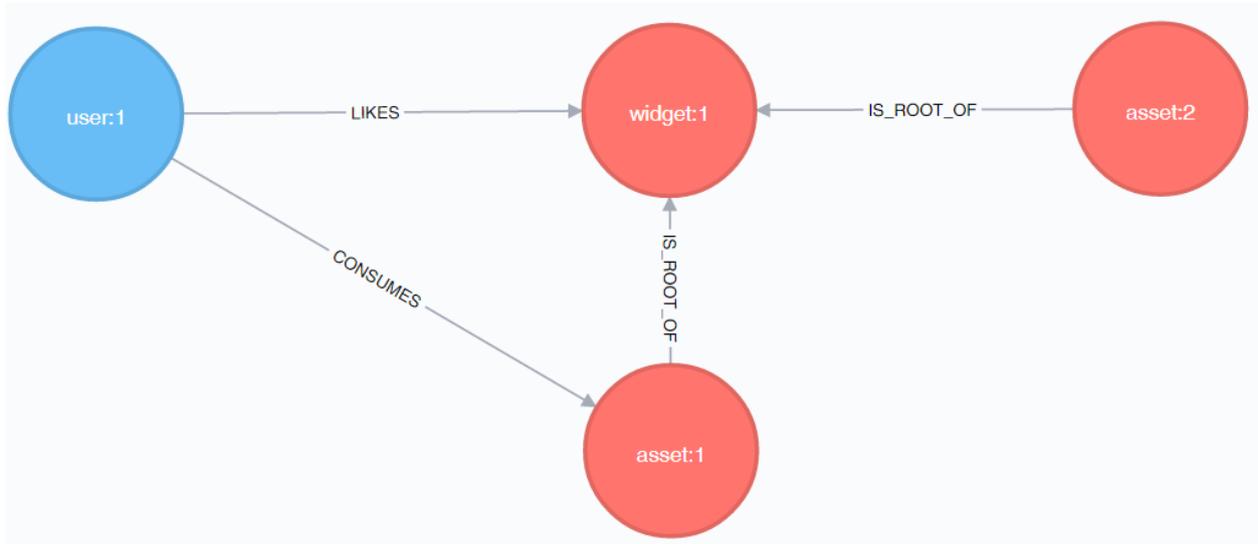

Figure 1. User interacts with widget.

When we are asked to rank widgets for a particular movie, we use interactions with widgets belonging to that movie (case 1) and interactions with movies that share widgets with the movie in question (case 2).

If a user has a direct interaction with an associated widget we use it 'as is' in our computations. We also consider indirect links, for example when a user "liked" a root asset, which shares widgets with the under investigation asset, we apply a weight of

$$W_a = \frac{r_x}{a+1}$$

where $r_x$ = rating of neighboring node, a = number of neighboring to the widget assets.

Therefore, the overall relevance weight of a person for a widget of an unconsumed asset becomes:

$$W = W_w + W_a = \sum r_x + \sum \frac{r_x}{a}$$

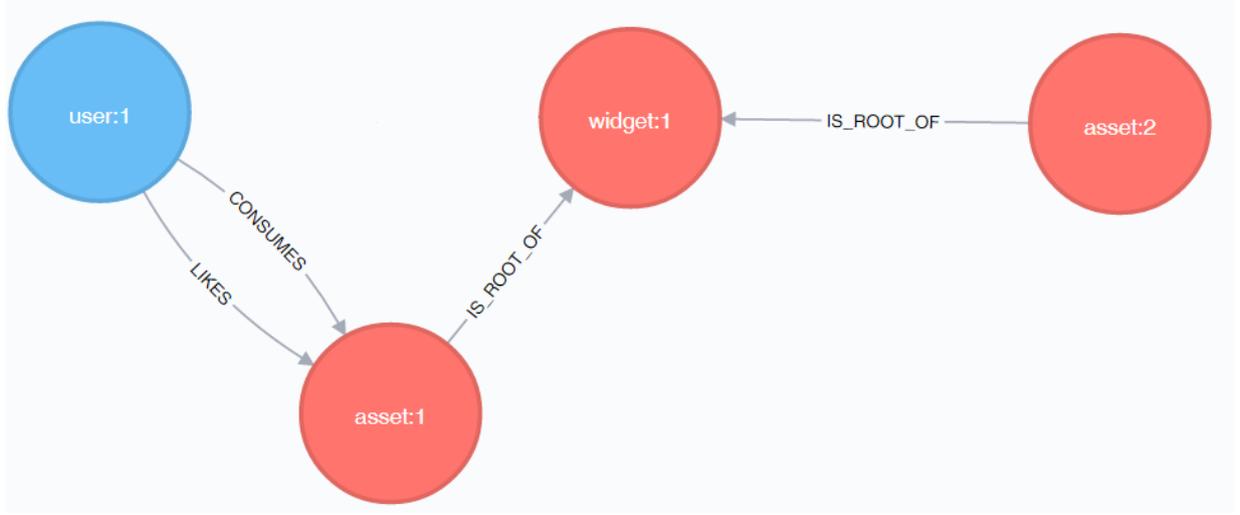

Figure 2. User interacts with movie

3.4. **Collaborative filtering analysis**

The technique described cannot provide rich results for assets that the user has not interacted with or with their neighbours ("isolated" assets). Thus, as a backup solution, collaborative filtering is applied among users, so as to estimate their relevance with such assets, based on her correlation with other users. A common approach for collaborative filtering having a dataset of simple numeric ratings [14], as in our evaluation dataset, is using the Pearson Correlation Coefficient. Its equation is the following:

$$c_{au} = \frac{\sum_{i=1}^{h}(r_{ai} - \bar{r}_a) \times (r_{ui} - \bar{r}_u)}{\sqrt{\sum_{i=1}^{h}(r_{ai} - \bar{r}_a)^2 \times \sum_{i=1}^{h}(r_{ui} - \bar{r}_u)^2}}$$

for users $a$ and $u$, where in our case $h = |I_{au}|$ is the amount of assets rated by both users, $r_{ai}$ is user $a$'s weight for asset $i$ and $\bar{r}_a = average(r_{a1}, r_{a2}, \ldots, r_{ah})$.

After calculating the correlation coefficients of a user with other users, Pearson collaborative filtering can provide a prediction, rating her relevance with an asset $j$, based on other users' relevance for the specific asset and their correlation:

$$p_{aj} = \bar{r}_a + \frac{\sum_{u=1}^{g}(r_{uj} - \bar{r}_u) \times c_{au}}{\sum_{u=1}^{g}|c_{au}|}$$

where $g$ is the number of users that consumed $j$ and $p_{aj}$ is the predicted rating of relevance for user $a$.

## 4. Experiments

### 4.1. Dataset and Configuration

Finding a dataset that contains user actions or in general two-level (implicit and explicit) data, proved challenging. In the end, we used a well known movie rating dataset found online [15], comprising a huge database of movies and user ratings, as well as keywords linked with those movies.

The dataset imported was interpreted into the SAM logic, directly importing SAM users, assets and keywords. We analysed the rating values in order to generate like/dislike actions based on those values (explicit information), and we were also able to use the implicit information of connected movies to the same keywords, thus having the two-level information that is necessary for our algorithm to display its full potential.

To limit our scale and make a meaningful analysis, we selected a random sample of available movies along with all the associated ratings and keywords. The overall numbers of the initial dataset imported can be found in the following table:

Table 2 – Sample retrieved from MovieLens dataset

| Sample metrics | |
|---|---|
| **Users** | 656 |
| **Movies** | 1032 |
| **Ratings** | 9902 |
| **Keywords** | 664 |

The dataset was split in a 70/30 ratio into a training and a testing set respectively. All experiments were performed on a desktop machine with an Intel Core TM i5-3400 Processor, 2.80 GHz, 12GB of RAM memory, running 64-bit Windows 10 Pro N.

### 4.2. Experiment Results

The graph analysis, supported by the Pearson Collaborative filtering, presented above, was applied and compared with the stand-alone implementation of Pearson Collaborative filtering as well as an implementation of k-nearest neighbours (K-NN) algorithm run over Neo4j. In Table 3 a report of errors for the testing set is presented.

Table 3 - Results of different algorithms run over Neo4j database.

| Algorithm | Mean absolute error | Root mean squared error | Mean percentage error |
|---|---|---|---|
| K-NN algorithm | 0.3415 | 0.4242 | 17.08% |
| Stand-alone Pearson Collaborative filtering | 0.2809 | 0.3190 | 14.04% |
| **SAM algorithm sup. by Pearson Collaborative filtering** | **0.2584** | **0.2878** | **12.92%** |

Based on measured errors, it is evident that the graph analysis is superior to the collaborative filtering approach when there is adequate user interactions. In case where there is not enough user interactions we fall back to Pearson's filtering. In addition, we can see that SAM algorithm outperforms K-NN, accuracy-wise, one of the most popular clustering approaches for recommendations using graphs.

Apart from the accuracy experiments, we also measure the response time of the three different algorithms (SAM, Collaborative filtering, K-NN) exposed by SAM's Context Management component as web services. Table 4 presents the average response time for each algorithm after 1000 requests on each. SAM algorithm's locality search seems to outperform other approaches.

Table 4 - Response Time Metrics

| Algorithm | Average Response Time (in ms) |
|---|---|
| K-NN algorithm | 6910 |
| Stand-alone Pearson Collaborative filtering | 5882 |
| **SAM algorithm sup. by Pearson Collaborative filtering** | **5529** |

## 5. Conclusions and Future work

This work has been focused on an efficient Context Management and Personalized Recommendation system for Social TV First and Second Screen. Second Screen Content Listening and related Recommendations is a new promising area that has yet to be explored by the research community. To this end, the authors have proposed an innovative and adaptive model, using social media and user context information, and applied over SAM's Content Syndication and Social learning environment. This model was supported by a collaborative filtering mechanism and evaluated over real-world dataset found online.

In the future, SAM will be piloted to schools and broadcasting agencies (Deutche Welle) as an eLearning and Media Delivery application, in order to test its functionalities and acquire real datasets of user interactions. Those interactions will constitute a more concrete dataset, to be used in order to evaluate the current representation model and the resulting recommendation system's effectiveness.

## 6. Acknowledgment

This work has been supported by the SAM project and funded from the European Union's 7th Framework Programme for research, technological development and demonstration under grant agreement no 611312.